\documentclass[final,5p,times,twocolumn]{elsarticle}
\usepackage{lineno,hyperref}
\usepackage{comment}
\usepackage{amsmath,amssymb}
\hypersetup{
    colorlinks=true,
    linkcolor=blue,
}
\usepackage[figurename=Fig., tablename=Table]{caption}
\usepackage{color}
\usepackage{cases}
\usepackage{ulem}
\usepackage{stfloats}
\usepackage{makeidx}
\modulolinenumbers[1]
\journal{Journal of \LaTeX\ Templates}
\biboptions{sort&compress}
\begin{document}
\begin{frontmatter}

\title{Analyzing the neutron and $\gamma$-ray emission properties of an americium-beryllium tagged neutron source}

\newcommand{\AFFicrr}{\address[AFFicrr]{Kamioka Observatory, Institute for Cosmic Ray Research, University of Tokyo, Kamioka, Gifu 506-1205, Japan}}
\newcommand{\AFFynu}{\address[AFFynu]{Department of Physics, Yokohama National University, Yokohama, Kanagawa, 240-8501, Japan}}
\newcommand{\AFFtus}{\address[AFFtus]{Department of Physics, Faculty of Science and Technology, Tokyo University of Science, Noda, Chiba 278-8510, Japan}}
\newcommand{\AFFokayama}{\address[AFFokayama]{Department of Physics, Okayama University, Okayama, 700-8530, Japan}}

\author[AFFtus]{Hiroshi~Ito\corref{CorrespondingAuthour}}
\cortext[CorrespondingAuthour]{Corresponding author}
\ead{itoh.hiroshi@rs.tus.ac.jp}
\author[AFFynu]{Kohei~Wada}
\author[AFFicrr]{Takatomi~Yano}
\author[AFFokayama]{Yota~Hino}
\author[AFFtus]{Yuga~Ommura}
\author[AFFokayama]{Masayuki~Harada}
\author[AFFynu]{Akihiro~Minamino}
\author[AFFtus]{and Masaki~Ishitsuka}

\AFFtus
\AFFynu
\AFFicrr
\AFFokayama

\begin{abstract}
Americium–beryllium (AmBe), a well-known tagged neutron source, is commonly used for evaluating the neutron detection efficiency of detectors used in ultralow background particle physics experiments, such as reactor neutrino and diffuse supernova neutrino background experiments. In particular, AmBe sources are used to calibrate neutron tagging by selecting the 4438-keV $\gamma$-ray signal, which is simultaneously emitted with a neutron signal. Therefore, analyzing the neutron and $\gamma$-ray emission properties of AmBe sources is crucial. In this study, we used the theoretical shape of a neutron energy spectrum, which was divided into three parts, to develop models of the energy spectrum and verify the results using experimental data. We used an AmBe source to measure the energy spectra of simultaneously emitted neutrons and $\gamma$-rays and determine the emission ratio of the neutrons with and without $\gamma$-ray emission. The measured spectrum was consistent with that obtained from the simulated result, whereas the measured emission ratio was significantly different from the corresponding simulated result. Here, we also discuss the feasibility of determining the neutron emission rates from the spectra divided into three
parts.
\end{abstract}

\begin{keyword}
Neutron tagging\sep
Americium-Beryllium\sep 
Gamma-ray detector\sep
Neutron detector\sep 
\end{keyword}
\end{frontmatter}

\section{Introduction}
\label{Introduction}

Americium–beryllium (AmBe) is a widely used radioactive neutron source, which can be tagged with the accompanying 4438~keV $\gamma$-ray emission. It is commonly used to evaluate the neutron detection efficiency of particle detectors employed in particle physics experiments. Several experiments conducted \cite{Eur.Phys.J.C81.775, DoubleChooz.2006vya, DayaBay.2018heb, Abe_2022, Borexino.2019gps, SK2021-SRN, Super-Kamiokande.2021the} to focus on reactor and astrophysical neutrinos, such as the diffuse supernova neutrino background search, involved the use of the neutron tagging technique for distinguishing electron antineutrino ($\bar{\nu}_e$) interactions from other neutrino interactions or their background events. Coincidental detection of positrons and neutrons (neutron tagging) is a suitable route for identifying the following charged current reaction: $\bar{\nu}_e + p \to e^+ + n$. Contrarily, an AmBe source exhibits several neutron emission channels, which lead to the emission of neutrons with different kinetic energies accompanied by $\gamma$-ray emission. Evaluating the fraction of these channels and neutron energies corresponding to an AmBe source is crucial for precisely measuring the neutron tagging efficiency.

AmBe is composed of $^{241}\rm{Am}$ and $^{9}\rm{Be}$ and emits neutrons according to the following chain reaction: ${}^{241}\rm{Am}\rightarrow{}^{237}\rm{Np}+\alpha$, ${}^{9}\rm{Be}+\alpha\rightarrow{}^{12}\rm{C}^{*}~(\rm{or}~{}^{12}\rm{C})+n$. In this process, the carbon nuclei can adopt several excited states, although only the first excited state of $^{12}$C emits $\gamma$-rays. The characteristics of the neutrons emitted by AmBe, i.e., the energy spectrum, flux, and emission ratio of neutrons and $\gamma$-rays, are correlated to the ratio of these excited states. The source size and Be density affect the neutron energy spectra below 2.5~MeV~\cite{Vijaya1973TheNS},
which corresponds to the second or higher excited states of $^{12}$C, probably owing to the different energy losses of the primary $\alpha$-particles. This difference can also be attributed to the method of production of the AmBe source.

To date, several theoretical and experimental studies have been reported on the energy spectra of neutrons emitted by an AmBe source as well as the correlation between the excited states of carbon nuclei~\cite{DEGUARRINI1971277, Vijaya1973TheNS, Lorch1973, GEIGER1975315}. Figure~\ref{figure.neutron.spectrum.sim} shows the recently reported energy spectral measurements~\cite{Marsh1995, SNOAmBe2020}. Most neutrons are paired with the first excited state ($n_{\rm{1st}}$) and ground state ($n_{\rm GND}$) of the $^{12}$C nucleus. The first excited state, with an energy of 4438~keV, relaxes to its ground state by immediately emitting a $\gamma$-ray photon with an energy of 4438~keV. Neutrons paired with the second excited state of carbon ($n_{\rm{2nd}}$) also significantly contribute to the energy spectra.  The second state, with an energy of 7654 keV, is the Hoyle state~\cite{1954ApJS....1..121H}, which is critical in nuclear astrophysics and decays by emitting three $\alpha$-particles with a probability of $\sim$99.6\%, and only $\sim$0.4\% of the state decays by emitting two $\gamma$-ray photons (energy: 3215 and 4438~keV)~\cite{PhysRevC.102.024320}. These decayed $\alpha$-particles subsequently interact with $^{9}$Be with a definite probability and simultaneously emit a $\gamma$-ray photon (energy: 4438~keV) and neutron. However, considering the ratio of the activity of $^{241}$Am to the neutron yield in our source, this interaction probability is negligible in our source. In addition to the neutrons paired with higher excited states, break-up neutrons, induced by following the reaction: ${^{9}\rm{Be}}+\alpha\to{^{9}\rm{Be}^{*}}+\alpha$, ${^{9}\rm{Be}^{*}} \to {^{8}\rm{Be}} + n$, contribute to the energy spectra appreciably below 1~MeV~\cite{Vijaya1973TheNS}. The kinetic energy of these neutrons is similar to or lower than that of the $n_{\rm{2nd}}$ neutrons and is well separated from those of $n_{\rm{1st}}$ and ($n_{\rm{GND}}$). Therefore, these neutrons have been combined and treated as $n_{\rm{2nd}}$ in the following discussion.

In recent studies on precise measurements of whole neutron energy spectra, $^{3}$He proportional and recoil-proton proportional counters, e.g., stilbene organic scintillators, have been employed \cite{Lorch1973, Marsh1995, SNOAmBe2020, ISO-8529-1}. The partial neutron spectrum of $n_{\rm 1st}$ is also measured by coincidental detection of 4438~keV $\gamma$-rays using the neutron time-of-flight method~\cite{Scherzinger2015, Geiger1964}, although the energy resolution is worse than those of the proportional counters. Because of the differences in the experimental setups and energy resolutions, distinguishing the $n_{\rm GND}$, $n_{\rm 1st}$, and $n_{\rm 2nd}$ spectra via a simple comparison of the experimental data is challenging. To circumvent this drawback, we introduced an experimental method to measure the ratio of $n_{\rm GND}$ and $n_{\rm 1st}$ directly using detector simulations and neutron spectral models. 

In this paper, we propose a model of neutron emission from an AmBe source and assess its validity by performing experimental measurements of the emission rates paired with each $^{12}\rm{C}^{*}$ level and the neutron energy spectrum accompanying $\gamma$-ray emission. 
Neutron energy spectral models, which are useful for simulating neutrons, combined with experimental evaluation of the characteristics of an individual AmBe source in a laboratory are crucial for realizing an efficient neutron tagging.
The remainder of this paper is organized as follows. 
The experimental details, including the characteristics of the AmBe source used in our study, components, simulation details, and experimental setups, are described in section~\ref{sec.MateMetho}. The analysis results, measured neutron and $\gamma$-ray energy spectra, and emission ratio of neutrons with/without $\gamma$-ray emission, are described in section~\ref{sec.discuss}. Finally, the major conclusions drawn from this study are elucidated in section~\ref{sec.conclusion}.

\section{Material and Methods}
\label{sec.MateMetho}

We performed two measurements for neutrons and $\gamma$-rays emitted from our AmBe source using NaI(Tl) crystal
and liquid scintillation detectors. The properties of the AmBe source were determined by comparing the
measured data and simulation results.

\subsection{Americium-Beryllium source}
\label{sec.AmBe}

The AmBe source, developed by the French Alternative Energies and Atomic Energy Commission in the early 1990s, was supplied by the Kamioka Observatory, Institute for Cosmic Ray Research, University of Tokyo, Japan. 
In this source, $^{241}$Am, with a strength of $97~\mu\rm{Ci}$ (3.6~MBq), is encapsulated with $^9\rm{Be}$
inside a stainless cylinder with the same outer diameter and height, i.e., 12.5~mm.
The neutron yield measured in March 2022 by the National Institute of Advanced Industrial Science and Technology (AIST), Japan, was $236.8\pm5.0 ~ n \cdot \rm s^{-1}$ ($S_n$). 
For the yield measurements, we compared the National Standard Field of Thermal Neutrons at AIST under the same conditions,
i.e., the AmBe (standard) source located in a graphite pile. 
The particles were counted using a $^3\rm{He}$ proportional counter~\cite{AIST2006}.

\begin{figure}[hbtp]
\centering
\includegraphics[width=0.45\textwidth]
{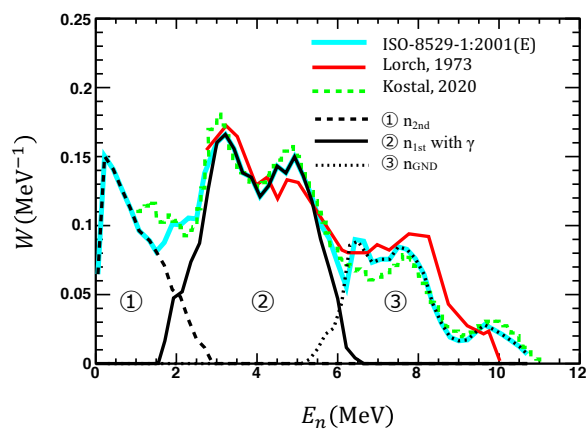}
\caption{
Measured neutron energy spectra of an AmBe source based on ISO 8529-1:2001(E)~\cite{Marsh1995, ISO-8529-1}, E. Lorch ~\cite{Lorch1973}, and M. Kostal et al.~\cite{SNOAmBe2020}. These measurements are conducted without considering the excited states of $^{12}$C. $n_{\rm{GND}}$, $n_{\rm{1st}}$, and $n_{\rm{2nd}}$ represent the partial neutron spectra with the following states of $^{12}$C: ground state, first excited state, and second or higher excited states, respectively.
}
\label{figure.neutron.spectrum.sim}
\end{figure}

Next, we performed a Monte Carlo simulation to analyze the experimental results, and the model of the
neutron energy spectra paired with the excited states of $^{12}$C is shown in Figure~\ref{figure.neutron.spectrum.sim}.
We adopted the International
Standard ISO-8529-1:2001(E) to model the total neutron spectra~\cite{ISO-8529-1}.
The spectra were divided into three parts
based on the theoretical predictions of De Guarrini and Malaroda~\cite{DEGUARRINI1971277}.
These partial neutron spectra were paired
with the states of $^{12}$C: ground state ($n_{\rm{GND}}$), first excited state ($n_{\rm{1st}}$), and the second excited state or higher states ($n_{\rm{2nd}}$).

\subsubsection{Measurement of $\gamma$ intensity from the AmBe source}

The intensity of the 4438 keV $\gamma$-rays was measured using a high-purity Ge (HPGe) detector (ORTEC, GEM110P4-ST), whose detection efficiency was evaluated via a Geant4-based Monte–Carlo simulation (also see section 2.3). The characteristics of the materials and inner structural dimensions of the HPGe detector and AmBe source were included in the simulation. The systematic uncertainty in the simulated $\gamma$-rays peak detection efficiency of the detector was $\sim$10\%, which was determined by comparing the simulated and calibration data of the detector for the 1173, 1333 (single $\gamma$-ray), and 2505~keV (sum of two $\gamma$-rays) peaks emitted by a $^{60}$Co source (19.5~kBq).

Figure~2~(a) and (b) show the energy spectra obtained from our AmBe source;
the spectra measurements were performed at distances of 10 and 50~mm from the surface of the HPGe detector, respectively.
The full widths at half maximum of the peak at 4438~keV are 119 ($L=10~\rm{mm}$) and 114~keV ($L=50~\rm{mm}$), which are consistent with previously reported values, obtained by considering Doppler broadening at the AmBe source~\cite{Doppler};
$\gamma$-ray peak broadening was considered in the simulation of the 4438~keV $\gamma$-ray spectrum for determining the detection efficiency of the detector. The energy resolution of the HPGe detector was negligible compared to the broadening caused by the Doppler effect.
In our case, the measured intensity of the 4438-keV $\gamma$-ray photons emitted from our AmBe source was $110.1\pm15.5~\gamma/\rm s$ ($S_\gamma$).

For our source, the emission ratio of the 4438-keV $\gamma$-ray photons to the neutron yield (${R_{\gamma/n} = S_\gamma/S_n}$, where $S_\gamma$ and $S_n$ indicate the emission rates of the 4438-keV $\gamma$-rays and neutrons from the AmBe source, respectively) was determined as $0.46\pm0.07$. This estimated ratio is consistent with those reported previously ($R_{\gamma/n} = 0.57$–-$0.75$)~\cite{LIU20071318} for AmBe sources within experimental errors. However, the observed differences between the ratio calculated in this study and those reported previously, albeit small, may reveal the characteristics of the corresponding AmBe sources used for these measurements.

\begin{figure}[h]
\centering
\includegraphics[width=0.45\textwidth]
{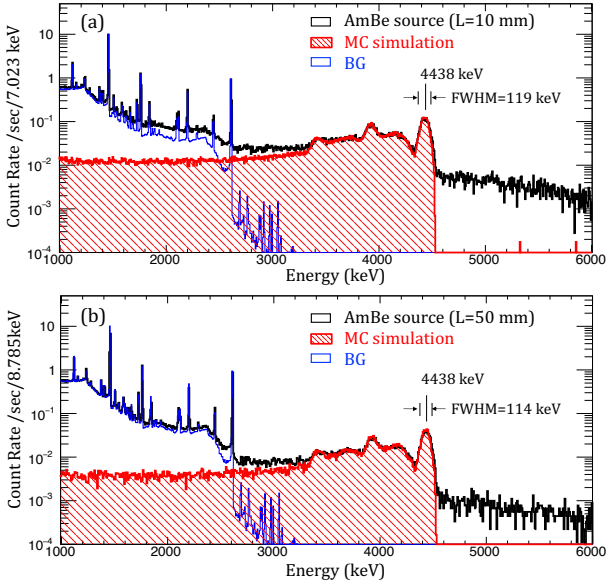}
\caption{
Energy spectrum of an AmBe source (black) and background (blue) measured using an HPGe detector, and the simulated spectrum of 4438-keV $\gamma$-rays emitted from the AmBe source (red). (a) 10~mm and (b) 50~mm distance between the center of the AmBe source and the HPGe surface}
\label{figure.energy.HPGe}
\end{figure}
    
\subsection{Equipments}

A NaI(Tl) crystal, with dimensions of $\rm 51~mm \times 51~mm \times 152~mm$, was optically connected to a two-inch photomultiplier tube. 
We used the assembled detector (OKEN, 8x8S24/1,5.VD(p), Oyo Koken Co. Ltd.) in this study. 
The crystal was then surrounded by 1-mm-thick aluminum plates for optical shielding.
The liquid scintillation detector consisted of a liquid scintillator (diameter: 3"; height: 3"; Saint-Gobain, BC501A)
connected to a three-inch photomultiplier tube.
The waveforms obtained from the NaI(Tl) and liquid scintillation detectors were stored using an oscilloscope (Lecroy, 100MXi) with a sampling rate of $\rm 1~GS~s^{-1}$.

The NaI(Tl) detector was used as a $\gamma$-ray energy spectroscope, whereas with the liquid scintillator detector, the pulse-shape discrimination (PSD) method was applied to separate the neutron and $\gamma$-ray signals.
The PSD and calibration methods are described in detail in Appendix~A. 
The calculated energy resolution was approximately 2\% in the main target energy region above 5~MeV, and the timing resolution was 1.72~ns. Based on the PSD of the pulses obtained from the liquid scintillation detector, the performance was confirmed to be $3\sigma$ for the neutron like event selection, and the $\gamma$-like contamination was less than 0.15\%.

\subsection{Monte Carlo simulation}
\label{sec.MC}

In this study, we used the GEANT4 Monte Carlo simulation tool~\cite{GEANT4} (Geant4.9.6p04 + G4NDL4.2) in order to evaluate the neutron and $\gamma$ emission of the AmBe source and take into account the detector response to the measurements.
In particular, neutron interactions were carefully treated because of the uncertainty in several models.

We prepared a geometrical setup for the simulation in which the source capsule was treated as a uniform stainless-steel cylinder to reproduce the size and total weight of our AmBe source. Further, the material characteristics and geometries of the NaI(Tl) and liquid scintillators were reproduced. In addition, to incorporate the contribution of scattered neutron background, the distance between the detector and the wall and the dimensions of the test bench (see section 2.4.1) used in the experiment were simulated.
Although the details of the detector, such as the inner parts of the photomultiplier tubes, were approximated, the uncertainty of the material composition was considered in the results.
To determine the energy resolution of our NaI(Tl) detector, a $\gamma$-ray source was used to irradiate the detector (for calibration), and the resulting output spectra of the detector were fitted with Gaussian functions. Further, the position of the $\gamma$-ray source was varied during the calibration process to determine the non-uniformity in the light yield of the detector, which was ultimately found to be negligible (see Appendix A.1). In addition, the timing resolution of the detector/experimental setup was evaluated for the subsequent neutron energy measurements, which were performed using the time-of-flight method (see Appendix A.2).

Fast neutrons with kinetic energies on the MeV scale mostly interact with the sodium or iodine nuclei in the NaI(Tl) scintillator.
In this study, we refer to the recent quenching factor reported by Joo et al.~\cite{Joo2019} as a function of the recoil energy for sodium and iodine.
 
In this study, we used models of the neutron energy spectra to simulate $\textcircled{2}$ $n_{\rm 1st}$ and $\textcircled{3}$ $n_{\rm GND}$ as shown in Fig.~\ref{figure.neutron.spectrum.sim}.
In particular, we validated the $n_{\rm 1st}$ spectral shape by considering the detector response. We considered only $n_{\rm GND}$ and n1st because the setups (explained in the next subsection) were designed to minimize the contribution of $n_{\rm 2nd}$; the simultaneous emission of $\gamma$-rays and $n_{\rm 2nd}$ is rare, and the amount of energy deposited in the NaI(Tl) scintillator is small.

\subsection{Setups}

The experimental setups used for measuring the neutron and $\gamma$-ray energy spectra are presented in this section, and the corresponding measurement results and their analyses are discussed in section 3.

\subsubsection{Energy spectrum of the neutron with $\gamma$ emission}
\label{sec.result.TOFene}

\begin{figure}[hbtp]
\centering
\includegraphics[width=0.45\textwidth]{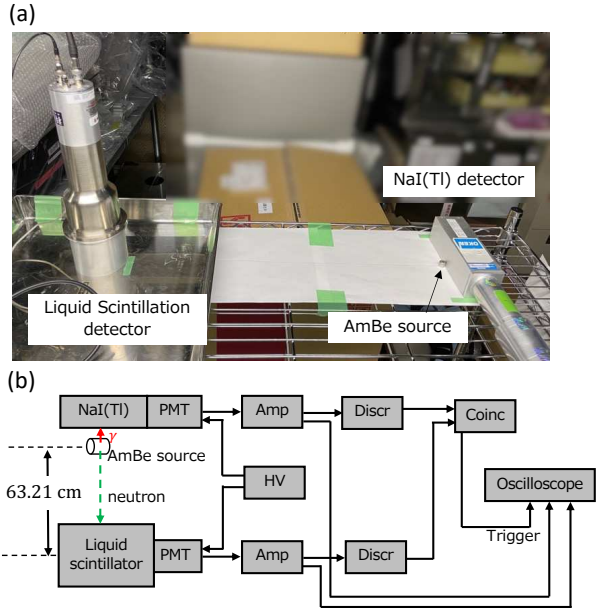}
\caption{
Schematics of the (a) experimental setup and (b) front-end data acquisition system used to measure the neutron energy spectrum based on the time-of-flight method.}
\label{figure.setup.b}
\end{figure}

To determine the kinetic energy of the neutrons, we used the time-of-flight method, in which the neutron kinetic energy was estimated from the difference between the arrival times of the neutrons and the $\gamma$-ray photons due to their different velocities.
The NaI(Tl) and liquid scintillation detectors were set at a distance from each other; the AmBe source was placed near the NaI(Tl) detector and 63.21~cm away from the center of the liquid scintillation detector as shown in Figure~\ref{figure.setup.b}.
Both the detectors and the source were placed on a stainless-mesh shelf (height: 120~cm) to reduce measurement uncertainties, which may have different origins; for example, the uncertainty caused by the event rate of the neutrons scattered by the surrounding materials (such errors were not observed in the simulation results). Furthermore, the detectors were set at distances of greater than 1~m from the walls and ceiling, and the coincidence signals were stored using an oscilloscope. The neutron emission events were determined by tagging the 4.4-MeV peak in the energy spectrum obtained from the NaI(Tl) detector. The neutron and $\gamma$-ray background signals were distinguished via PSD using the
 liquid scintillation detector.

\subsubsection{Determination of event ratio of higher energy tail to 4.4~MeV peak}
\label{sec.result.Rn0n1}

\begin{figure}[hbtp]
\centering
\includegraphics[width=0.47\textwidth]{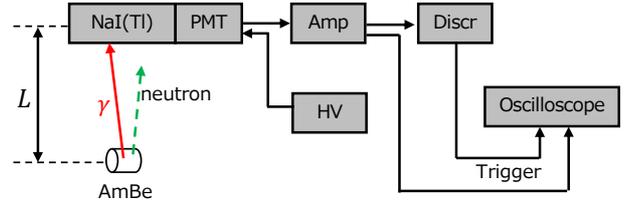}
\caption{
Schematic of the experimental setup and front-end data acquisition system used to determine the neutron emission ratio of $n_{\rm GND}$ to $n_{\rm 1st}$ with $\gamma$-ray emission by varying the distance $L$.}
\label{figure.setup.c}
\end{figure}

We determined the neutron emission ratio of $n_{\rm GND}$ to $n_{\rm 1st}$ with $\gamma$-ray emission ($R_{n0/n1}$) from the AmBe source using the setup depicted in Figure~\ref{figure.setup.c}. To measure the energy spectra using the NaI(Tl) detector, the distance ($L$) between the centers of the AmBe source and the NaI(Tl) detector was varied to 3.5, 4.5, 5.5, 8.5, 10.0, 13.5, 16.0, 18.5, 28.5, and 38.5~cm. The ratio of the number of events in the high-energy tail components (5--11~MeV) of the 4.4~MeV peak was measured. The experiments were performed in the energy region of 3--10~MeV to ensure that the neutrons of the second or higher excited states remain almost unaffected. The emission ratio of the neutrons with and without $\gamma$-ray emission was determined by comparing the observed and simulated event ratios of the peak and tail components.

\section{Result and Discussion}
\label{sec.discuss}

\subsection{Neutron with $\gamma$-ray energy spectra measured using the time-of-flight method}

\begin{figure}[h]
\centering
\includegraphics[width=0.49\textwidth]{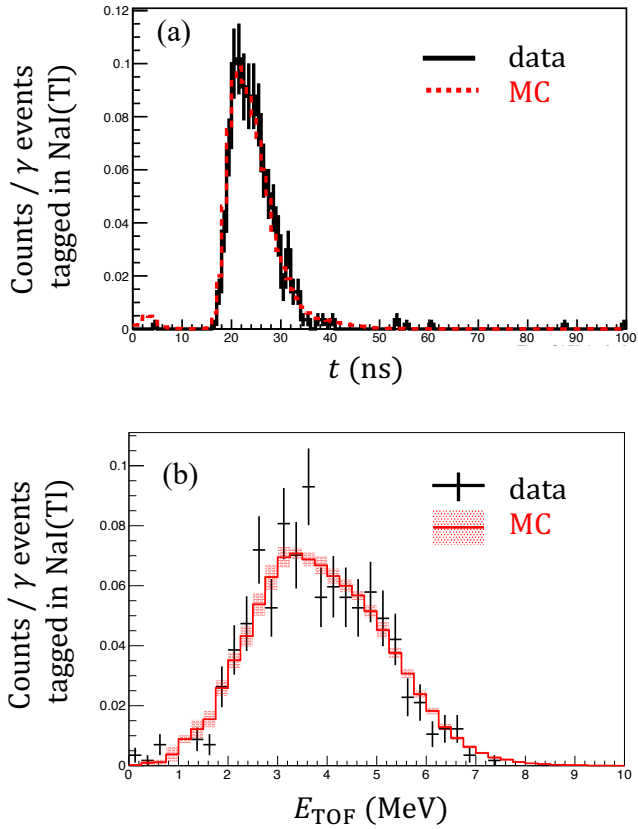}
\caption{
Arrival time distribution (a) and neutron energy spectra of data and simulation based on time-of-flight (b). Both events are selected by requiring neutron identified by PSD of the liquid scintillation detector and $\gamma$ detection in the NaI(Tl) detector. The black dots and red lines are data and the simulation, respectively.}
\label{fig:neutron.spectrum}
\end{figure}

The neutron-like events were discriminated by analyzing the pulse shapes of the signals obtained from the liquid scintillator. The arrival-time distribution is shown in Figure~\ref{fig:neutron.spectrum} (a). The black solid and red dashed histograms represent the experimental data and the simulation based on the ISO model with smearing by the measured timing resolution, respectively. Then, based on the time of flight, the neutron kinetic energy is converted from the arrival time difference between the NaI(Tl) and liquid scintillation detectors, given as
\begin{eqnarray}
    E_{\rm TOF} = m_n\left( \frac{1}{\sqrt{1-(L/ct)^2}} - 1 \right),
\end{eqnarray}
where $m_{\rm n}$ is the neutron mass, $L$ is the path length of the neutron (63.21~cm), and $ct$ is the path length of light during flight.

The measured neutron energy spectra and the expectations from the simulation are shown in Figure~\ref{fig:neutron.spectrum}~(b). The simulation band indicates systematic uncertainty, which is dominated by the model difference of the cross-section of neutron scattering with iodine (Appendix~B). In comparison with the simulation, $\chi^2/dof=28.7/40$ is calculated, which corresponds to the p-value of 0.908; this value indicates that the measured energy spectrum is in good agreement with that simulated using the prediction model (shown in Figure~\ref{figure.neutron.spectrum.sim}).

\begin{figure*}[!h]
\centering
\includegraphics[width=0.9\textwidth]
{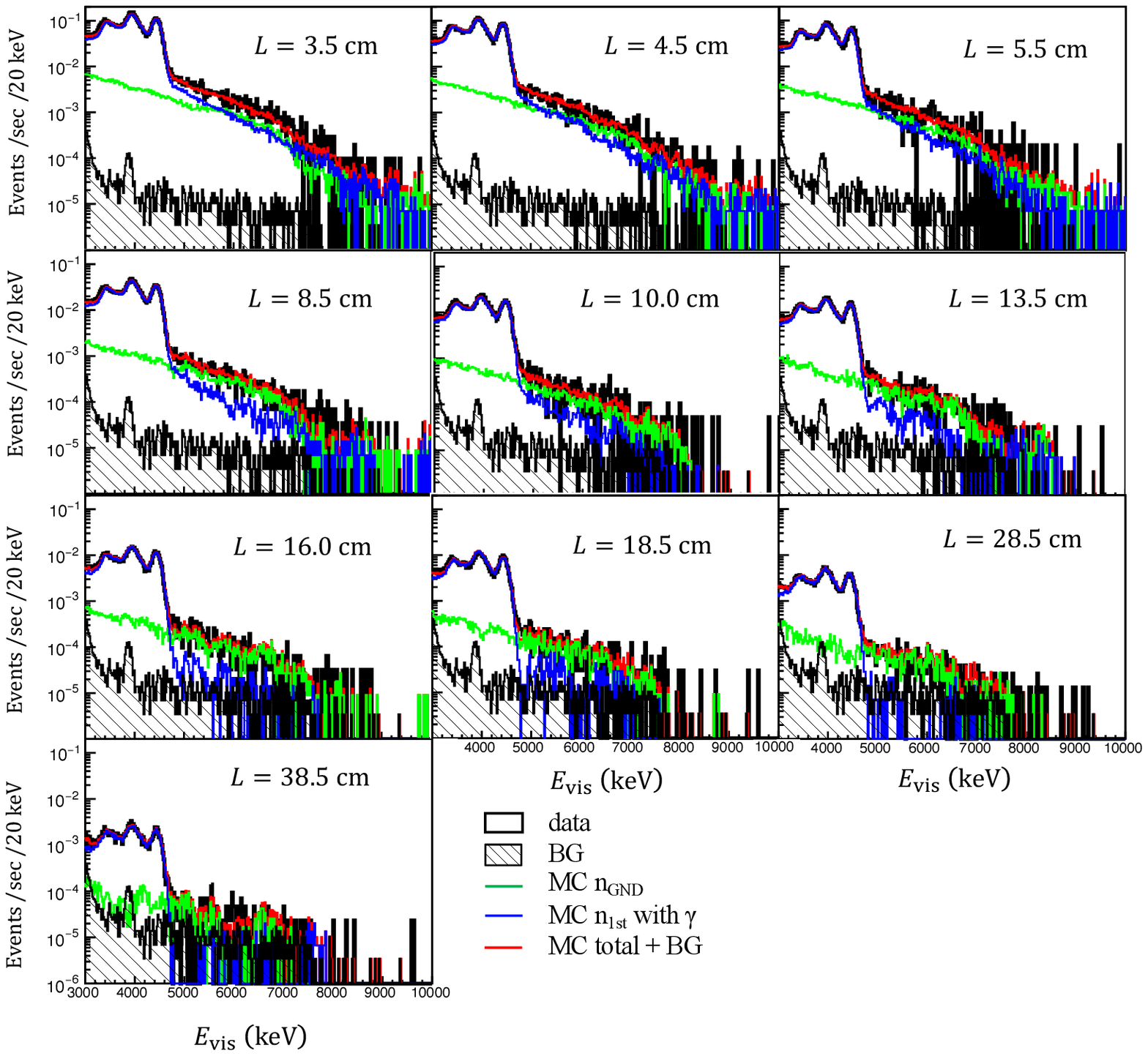}
\caption{
Distribution of visible energy in the NaI(Tl) detector for a distance between the center of the AmBe source and the NaI(Tl) detector ($L$). Data and BG represent measured spectra with the AmBe source and without the source. MC $n_{\rm GND}$ and MC $n_{\rm 1st}$ indicate simulated spectra for neutrons with the ground state and 1st exited state of carbon. In a case of 1st exited state, 4438 keV $\gamma$ ray also emitted simultaneously.}
\label{figure.ratio.hist}
\end{figure*}

\subsection{Emission ratio of neutron with/without $\gamma$-ray emission }

The distribution of the visible energy for each distance $L$ is shown in Figure~\ref{figure.ratio.hist}.
The high-energy tail beyond the 4438~keV $\gamma$-ray peak can be attributed to the neutrons emitted from the source.
A fraction of the emitted neutrons interact with the nuclei in the NaI(Tl) detector and excite them via inelastic scattering.
The $\gamma$-rays emitted by these nuclei upon deexcitation are detected by the detector via the detection of energy-deposition-induced scintillations in the scintillator medium: 
e.g., $n+{^{22}\rm Na}\to n+{^{22}\rm Na}^*$, $^{22}{\rm Na}^*\to{^{22}\rm Na}+ \gamma(\rm s)$.
The nuclei may also recoil without undergoing excitation upon interacting with the incident neutrons. 
However, the energy deposited during this elastic scattering process (nuclear recoil) is smaller by an order of magnitude. 
This neutron inelastic scattering process is utilized to evaluate the emission ratio of $n_{\rm GND}$ and $n_{\rm 1st}$ by considering the different acceptances of the detector for a single neutron, pair of neutrons, and the 4438~keV $\gamma$-rays. 
These acceptances are approximately described as follows:proportional to  $L^{-2}$, $L^{-2}$, and $L^{-4}$
for a single $\gamma$-ray (4438 keV peak), single neutron (${E_{\rm vis}>5~{\rm MeV}}$), and coincidentally detected $\gamma$-ray and neutron (${E_{\rm vis}>5~{\rm MeV}}$), respectively.

The ratio of the high-energy tail (${E_{\rm vis}>5~{\rm MeV}}$) to the 4438~keV peak, $R_{\rm tail/peak}$, is calculated for each value of distance ($L$) between the source and the NaI(Tl) detector (see Figure~\ref{figure.ratio.hist}) using the following equation:
\begin{eqnarray}
R_{\rm tail/peak}(L) = 
\frac{
\int^{10,000~\rm keV}_{4,750~\rm keV} \Bigl\{ N(E_{\rm vis}, L)-N_{\rm BG}(E_{\rm vis}) \Bigr\} ~dE_{\rm vis}}
{\int^{4,536~\rm keV}_{4,321~\rm keV} \Bigl\{ N(E_{\rm vis}, L) - N_{\rm BG}(E_{\rm vis}) \Bigr\} ~dE_{\rm vis}},
\end{eqnarray}
where $N(E_{\rm vis}, L)$ is the event rate for distance $L$ and $E_{\rm vis}$ bin, and $N_{\rm BG}(E_{\rm vis})$ are the background rates for the $E_{\rm vis}$ bin.

\begin{figure}[!h]
\centering
\includegraphics[width=0.49\textwidth]
{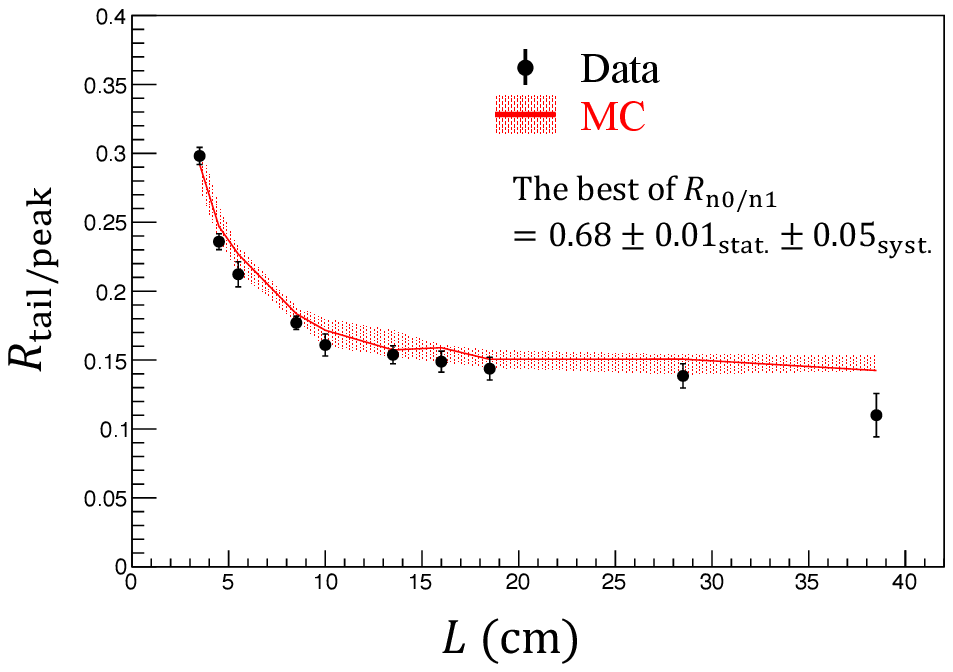}
\caption{
Correlation between distance $L$ and $R_{\rm tail/peak}$ used to determine the neutron emission ratio of $n_{\rm GND}$ to $n_{\rm 1st}$ with $\gamma$-ray emission. Black dots are the measured data, and the red line is the best fit simulated data. The red band shows a total systematic uncertainty, where a difference of cross-section models for neutron interaction is a dominant contribution in this study.}
\label{figure.ratio.plot}
\end{figure}

The ratio ($R_{\rm tail/peak}$) as a function of $L$ is shown in Figure~\ref{figure.ratio.plot}, which indicates the simulated $R_{\rm tail/peak}$ also varies with $L$ because of the different $L$ dependences of the peak and tail components.
As a result, we determined 
$R_{\rm n0/n1}$
to be $0.68\pm0.01_{\rm stat.}\pm0.05_{\rm syst.}$ by fitting the $R_{\rm tail/peak}$ distribution from the simulation to the data. The systematic uncertainty was dominated by the model difference in the cross-section of neutron scattering with iodine (Appendix~B). 
The best fit value of Rn0/n1 was determined with a minimum $\chi^2$ of $\chi^2$ of $\chi^2/dof=9.46/9$.

\subsection{Comprehensive profile of the AmBe source}

The measured emission parameters of neutrons and $\gamma$ are summarized in Table~\ref{table.result}. Based on this information, we deduced the emission rates of neutrons from the AmBe source, that is, $n_{\rm 1st}$, $n_{\rm GND}$, and $n_{\rm 2nd}$.

\begin{table*}[h]
\centering
\caption{Summary of measured emission
rate and ratio for the AmBe source.}
\begin{tabular}{l l}
\hline
\hline
Neutron yield & $S_{\rm n}=236.8 \pm 5.0 ~ n \cdot {\rm s^{-1}}$ \\
Intensity of 4.4 MeV $\gamma$ emission & 
$S_{\rm \gamma}=110.1\pm15.5~\gamma \cdot \rm s^{-1}$ \\
Emission ratio of $n_{\rm GND}$ to $n_{\rm 1st}$ & $R_{\rm n0/n1}=0.68 \pm 0.01_{\rm stat.} \pm 0.05_{\rm syst.}$ \\
\hline
\hline
\end{tabular}
\label{table.result}
\end{table*}

The $n_{\rm 1st}$ emission rate is assumed to be equal to $S_\gamma$. Second, the $n_{\rm GND}$ emission rate is calculated by using $R_{\rm n0/n1} \cdot S_{\rm \gamma}$ and is found to be ${74.9\pm11.9 ~ n \cdot {\rm s^{-1}}}$. Finally, the $n_{\rm 2nd}$ emission rate is estimated by subtracting $n_{\rm GND}$ and $n_{\rm 1st}$ from the total neutron yield and found to be ${51.8\pm17.2 ~ n \cdot {\rm s^{-1}}}$.
The results are consistent with the expectations for the $n_{\rm GND}$
and $n_{\rm 2nd}$ emission rates of $59.0\pm1.3~n \cdot {\rm s^{-1}}$ ($n_{\rm GND}$) and $54.2\pm1.1~n \cdot {\rm s^{-1}}$ ($n_{\rm 2nd}$), respectively. These expectations are given by
 our neutron energy spectrum models shown in Figure~\ref{figure.neutron.spectrum.sim} and the measured total neutron and 4438~keV $\gamma$ emission.
 As described above, our method can be used to determine the emission rates of $n_{\rm GND}$, $n_{\rm 1st}$, and $n_{\rm 2nd}$.

\section{Conclusion}
\label{sec.conclusion}

AmBe, which emits neutrons accompanied by $\gamma$-rays, is widely used to calibrate neutron detectors. We examined the energy spectra of the neutrons and $\gamma$-rays emitted from our AmBe source using NaI(Tl) and liquid scintillation detectors.
We modeled the neutron energy spectra paired with the respective excited states of carbon, divided into three parts, by referring to the theoretical prediction. 
The energy spectrum of $n_{\rm 1st}$ was evaluated based on the time of flight, which was consistent with our prediction considering its neutron energy spectrum and our detector response. The emission ratio of neutrons with/without $\gamma$ emission ($R_{n0/n1}$) was measured based on the calorimetric method with the NaI(Tl) detector by changing the distance between the source and the detector.
It was determined to be $R_{n0/n1} = 0.68 \pm 0.01_{\rm stat.} \pm 0.05_{\rm syst.}$
This method was established for
the first time. The $\gamma$ intensity of $110.1 \pm 15.5~ \gamma \cdot \rm s^{-1}$
from the AmBe source with a neutron fluence of $236.8 \pm 5.0 ~n \cdot \rm s^{-1}$ was obtained by performing independent measurements. The emission rates of $n_{\rm GND}$ and $n_{\rm 2nd}$ were determined to
be $74.9 \pm 11.9$ and $51.8 \pm 17.2 ~n \cdot \rm s^{-1}$, respectively; these are reasonably consistent with the expectations of our model.
Therefore, this method comprehensively determined the neutron emissivity of the excited states of Carbon in the AmBe source.
Our neutron spectral model showed good consistency in this study. 
The characteristics of the AmBe source can be understood by integrating the experimental measurement of the emission ratio of neutrons and the modeled energy spectra. Thus, these will contribute to the general use of AmBe sources, such as in measurements and calibrations.

\section*{Acknowledgments}

This work was partially supported by the joint research program of the Institute for Cosmic Ray Research (ICRR) at The University of Tokyo. This work was supported by the Neutron Measurement Consortium for Underground Physics, JSPS KAKENHI Grant Grants, Grant-in-Aid for Scientific Research (C) [grant number 20K03998], and Grant-in-Aid for Scientific Research on Innovative Areas [grant number 19H05808].

\section*{Appendix A. Calibration and performance check of the detectors}

\subsection*{A.1. Energy calibration of NaI(Tl) detector}
\label{sec.calib}

We calibrated the energy of the NaI(Tl) detector using $\gamma$ emission sources, $^{22}$Na, $^{137}$Cs, and $^{60}$Co. 
In addition, the environmental $\gamma$ sources $^{40}$K and $^{208}$Tl were used for the energy calibration. The relationship between the visible energy $E_{\rm vis}$ (MeV) and the integrated charge $q$ of the signal waveform (pC) was determined as a function of the energy as follows:
\begin{eqnarray}
q(E_{\rm vis}) = a \biggl[ 1 - \exp\left(-E_{\rm vis}/b \right) \biggr],
\end{eqnarray}
where $a$ and $b$ are the fit parameters. A saturation effect for the phototubes was found at approximately a few MeV, and a fit function with an exponential function was employed to account for this effect.

    \begin{figure}[hbtp]
    \centering
    \includegraphics[width=0.49\textwidth]
    {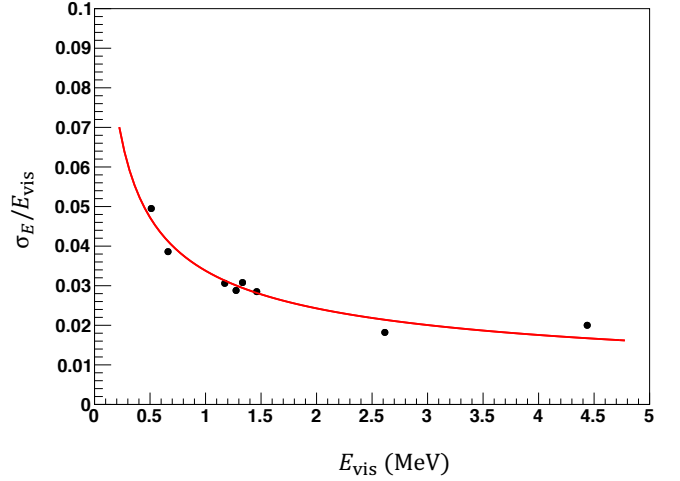}
    \caption{
    Energy resolution of the NaI(Tl) detector as a function of visible energy with a fit line.
    }
    \label{figure.energy.resolution}
    \end{figure}

Then, we measured the energy resolution ($\sigma_E(E_{\rm vis})/E_{\rm vis}$) of the detectors given as
\begin{eqnarray}
\sigma_E(E_{\rm vis})/E_{\rm vis} = \sigma_0 + \frac{\sigma_1}{\sqrt{E_{\rm vis}}}
\label{eq.ene.resolution}
\end{eqnarray}
where $\sigma_0$ and $\sigma_1$ are fit parameters.
The energy dependence of resolution $\sigma_E/E_{\rm vis}$ follows the function of Eq.~\ref{eq.ene.resolution}, as shown in Figure~\ref{figure.energy.resolution}, and the function with the best-fit parameters was used to reproduce the detector response in the simulation, as explained in Section~\ref{sec.MC}.

The dependence of the light yield’s uniformity on the position of the $\gamma$-ray source was investigated by measuring the integrated charge of the 662~keV $\gamma$-rays at different positions of a $^{137}$Cs source. The light yield was found to be uniform within an error of 0.4\%, and error propagation analysis showed that this error is negligible compared to the experimental energy resolution (Figure~\ref{figure.energy.resolution}).

\subsection*{A.2. Timing resolution of the NaI(Tl) and liquid scintillation detectors}

To measure the timing resolution of the setup as shown in Fig.~\ref{figure.setup.b} using two $\gamma$’s from annihilation, a $^{22}$Na source was placed in the center between two detectors, at a distance of 30~cm from each detector. Events in the 511~keV peaks were selected, and the time resolution and offset between the two detectors were determined based on the time difference.

In this study, the time walk effect (pulse height dependence of arrival time) was considered using the constant fraction discrimination (CFD) method in the waveform analysis. We optimized the CFD delay times for the signals of the NaI(Tl) and liquid scintillation detectors to 10 and 2~ns, respectively.

The timing resolution was measured to be 1.2~ns, and this value was used to smear the arrival time in the simulation to compare the measurements of neutrons and $\gamma$’s from the AmBe source. The possible bias was estimated to be 0.03~ns to determine the time offset and is negligibly small for the conversion of the time of flight of neutrons to kinetic energy.

    \begin{figure*}[h]
    \centering
    \includegraphics[width=0.9\textwidth]
    {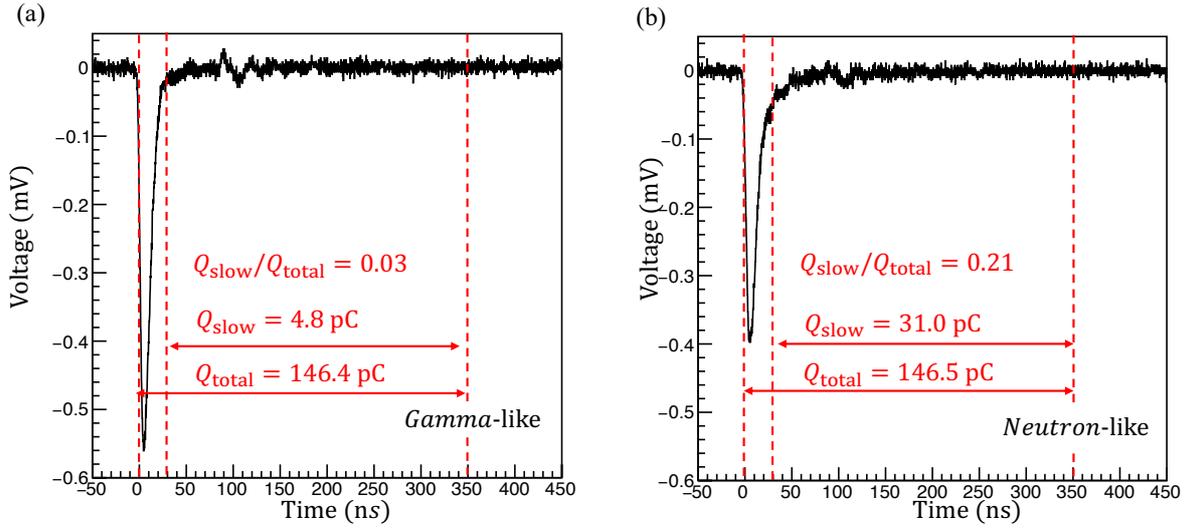}
    \caption{
    Typical waveforms of (a) $\gamma$-like and (b) neutron-like signals.
    }
    \label{figure.typical_wave_form_LSPSD}
    \end{figure*}
    
    %
    %
    \begin{figure}[h]
    \centering
    \includegraphics[width=0.49\textwidth]
    {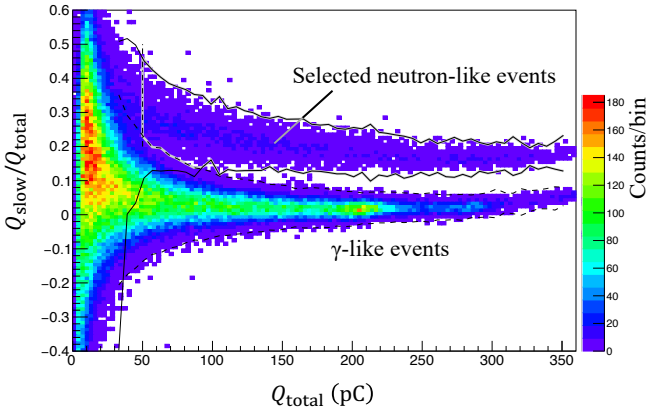}
    \caption{
    Relation between $Q_{total}$ and $Q_{\rm slow}/Q_{\rm total}$.
    Two components appear
    and black solid and dashed lines include 3$\sigma$ of neutron-like and $\gamma$-like events, respectively.
    In this study,
    as the neutron-like selection,
    the PSD selection region is within the gray solid line.
    }
    \label{figure.ngamma.discrimination}
    \end{figure}

\subsection*{A.3. Discrimination of neutron and $\gamma$}
\label{sec.PSD}

To discriminate neutron events from $\gamma$ events, the $^{252}$Cf source was placed at a distance of 63.21~cm from the liquid scintillation detector. Neutron and $\gamma$ events are discriminated by using the signal waveform, called PSD, because neutron events tend to have light emission in late timing (tail in the signal waveform), whereas the waveform of $\gamma$ events tends to be sharp.
Figure~\ref{figure.typical_wave_form_LSPSD} shows the typical waveforms of $\gamma$-like and neutron-like signals. 
In particular, variable $Q_{\rm slow}/Q_{\rm total}$ is defined as the ratio of the integrated charges of a signal in a liquid scintillation detector with different time windows, where $Q_{\rm slow}$ and $Q_{\rm total}$ are respectively given as
\begin{eqnarray}
Q_{\rm slow}=\int_{t_{\rm r} + 30~\rm ns}^{t_{\rm r}+t_{\rm w}} \frac{V(t)}{z}~dt,
\label{eq.ngamma2}
\end{eqnarray}
\begin{eqnarray}
Q_{\rm total}=\int_{t_{\rm r}}^{t_{\rm r}+t_{\rm w}} \frac{V(t)}{z}~dt,
\label{eq.ngamma3}
\end{eqnarray}
$V(t)$ is the amplitude of the waveform (mV) of the signal as a function of time t in a unit of ns), and ${z=50~\Omega}$ is the impedance of the cables. In the integral region, $t_{\rm r}$ and $t_{\rm w}$ are the rising times satisfying $V(t_{\rm r}) = V_{\rm max}/3$ and the gate time of the window, which was set to 350~ns in this study.

{Figure~\ref{figure.ngamma.discrimination}} shows the relationship between $Q_{\rm total}$ and $Q_{\rm slow}/Q_{\rm total}$, where two clear components appear at $Q_{\rm total} > 50~\rm pC$. The criteria are determined to select neutron-like events with $3\sigma$ range of $Q_{\rm slow}/Q_{\rm total}$ distribution sliced into $Q_{\rm total}$ in a region of $50 < Q_{\rm total} < 300~\rm pC$. The $\gamma$-like contamination in any $Q_{\rm total}$ is estimated to be less than 0.15\% in the selected region. We calibrated the conversion factor from energy deposition to $Q_{\rm total}$ of $96~\rm pC/MeV_{ee}$.

\section*{Appendix B. Systematic uncertainties}
\label{sec.app.b}

We considered the following systematic uncertainties: The main contribution is the difference in the physical model of neutron interaction with the material. Other sources that are estimated to have negligible effects are as follows: (1) the uncertainty of material components, such as the inner parts of photomultiplier tubes and the stainless mesh shelf, (2) the effect of bias to determine the timing offset for the NaI(Tl) and the liquid scintillation detector, and (3) the quenching factor uncertainty of nuclear recoil energy in the NaI(Tl) scintillator.

\subsection*{B.1. Neutron interaction with the material}

\begin{figure}[h]
    \centering
    \includegraphics[width=0.49\textwidth]
    {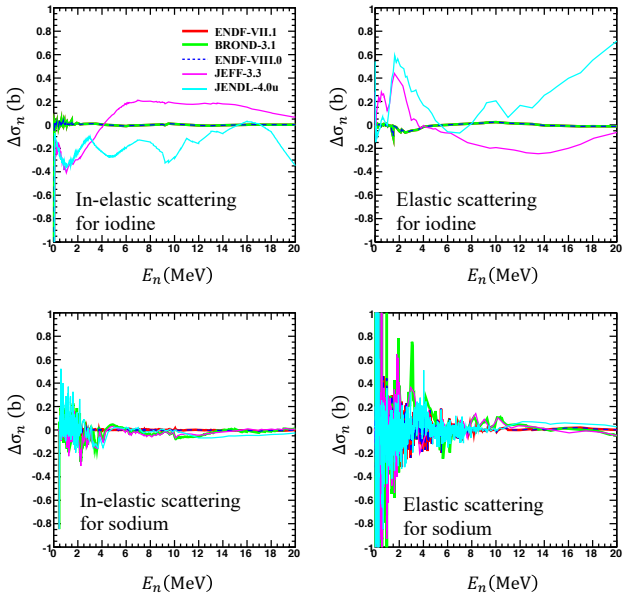}
    \caption{
    Difference of cross section from ENDF-VII.0 model as a function of neutron energy for elastic/inelastic scattering with iodine and sodium nuclei.
    Red, green, blue-dashed, magenta, and cyan lines are  ENDF-VII.1~\cite{ENDF2006}, BROND-3.1~\cite{BROND2016}, ENDF-VIII.0~\cite{ENDF2006}, JEFF-3.3~\cite{JEFF2020}, and JENDL-4.0u~\cite{JENDL2017} models, respectively.
    }
    \label{figure.cross.section}
    \end{figure}

We adopted the physical models of thermal and fast neutron interactions with the materials listed in Table~\ref{table.phys.list}.
These models are included in the default GEANT4 package and are reasonably consistent with typical fast neutron experiments, for example, the cross-section of elastic/inelastic scattering and nuclei capture. The ENDF-VII.0 model is adopted as a data set of the cross-section for neutrons~\cite{ENDF2006}, whereas the cross-sections for elastic and inelastic neutrons with 23Na and 127 311 I in the BROND~\cite{BROND2016}, JEFF~\cite{JEFF2020}, and JENDL~\cite{JENDL2017} models, which are often used in simulations, show some discrepancies from our data. Figure~\ref{figure.cross.section} shows the difference between cross-sections $\Delta \sigma_n$ and those obtained using the ENDF-VII.0 model as a function of neutron energy for elastic/inelastic scattering with iodine and sodium nuclei. We consider the difference between the models in the simulation to be a systematic uncertainty.

    \begin{table}[htbp]
    \centering
    \caption{Physics model of neutron interaction}
    \begin{tabular}{c c }
    \hline
    \hline
    Thermal neutron & G4NeutronHPThermalScattering\\
    ($<4$~eV)&\\
    \hline
    Fast neutron & G4NeutronHPorLElastic\\
    ($4$~eV $\sim$ 20~{\rm MeV}) & G4NeutronHPorLEInelastic\\
     & G4NeutronHPorLCapture\\
    \hline
    Data set & ENDF-VII.0~\cite{ENDF2006}\\
    \hline
    \hline
    \end{tabular}
    \label{table.phys.list}
    \end{table}

\bibliography{Reference.bib}

\end{document}